\documentclass[preprint]{aastex}
\usepackage{graphicx}
\usepackage{subfigure}
\usepackage {natbib}

\begin{document}

\title  { TWO EXTRASOLAR ASTEROIDS WITH LOW VOLATILE-ELEMENT MASS FRACTIONS}

\author{M. Jura\altaffilmark{a}, S. Xu(\includegraphics[width=1.2cm]{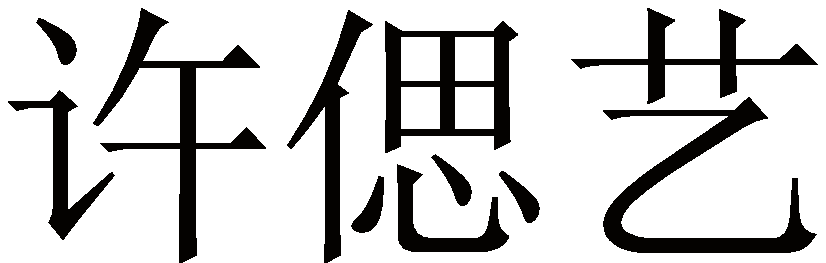})\altaffilmark{a}, B. Klein\altaffilmark{a}, D. Koester\altaffilmark{b} \& B. Zuckerman\altaffilmark{a}} 

\altaffiltext{a}{Department of Physics and Astronomy, University of California, Los Angeles CA 90095-1562; jura, sxu, kleinb, ben@astro.ucla.edu}

\altaffiltext{b}{Institut fur Theoretische Physik und Astrophysik, University of Kiel, 24098 Kiel, Germany; koester@astrophysik.uni-kiel.de}

\begin{abstract}
Using ultraviolet spectra obtained  with the {\it Cosmic Origins Spectrograph} on the  {\it Hubble Space Telescope}, we extend our previous ground-based optical determinations of  the composition of  the extrasolar asteroids accreted onto two  white dwarfs, GD 40 and G241-6.  Combining optical and ultraviolet spectra of these stars with He-dominated atmospheres, 13 and 12 polluting elements are confidently detected in GD 40 and G241-6, respectively.    For the material accreted onto GD 40,  the volatile elements  C and S  are  deficient by more than a  factor of 10 and N by at least a factor of 5 compared to their mass fractions 
in primitive CI chondrites and approach what is inferred for  bulk Earth.   A similar pattern is found for G241-6 except that S is undepleted. We have also newly detected or placed meaningful upper limits for the amount of Cl, Al, P, Ni and Cu in 
the accreted matter.    Extending results from optical studies, the mass fractions of  refractory elements in the accreted parent bodies are  similar to what is measured for bulk Earth and chondrites.  Thermal
processing, perhaps interior to a snow line, appears to be of central importance in determining the elemental compositions
of these particular extrasolar asteroids.  
\end{abstract}
\keywords{planetary systems -- stars, white dwarf}
\section{INTRODUCTION}
Evidence is now compelling that most externally-polluted white dwarfs  
derive their heavy elements\footnote{To avoid the specific connotation of ``metals" in the context of planetary physics, we use ``heavy elements"  to mean all elements heavier than helium.} by accretion from tidally-disrupted planetesimals \citep{Debes2002,Jura2003, Gaensicke2006, Jura2008, Farihi2010, Zuckerman2010}.  Here, we report new ultraviolet spectroscopy of two polluted white dwarfs to exploit this
 uniquely powerful tool to determine the bulk composition of extrasolar asteroids -- the building blocks of rocky planets.
 
The composition of extrasolar  planets is of great interest and considerable effort has been devoted to studying main-sequence stars with the goal of
 identifying spectroscopic signatures of accretion of planets  \citep{Ramirez2010, Gonzalezhernandez2011, Brugamyer2011, Schuler2011}.
 However, because a star's own heavy elements dominate the composition of its photosphere, it is very difficult to determine a planetary contribution to the measured abundances.
 In contrast,  as reviewed by \citet{Koester2009}, white dwarfs with effective temperatures less than 25,000 K typically have atmospheres that  are nearly pure hydrogen or pure helium  and the observed heavy elements usually are the result
 of external pollution\footnote{Exceptions include the DQ white dwarfs where carbon is dredged-up from the interior \citep{Koester1982,Pelletier1986}, ``warm" DQ stars where carbon  dominates the composition of the  photosphere   \citep{Dufour2007}, and  high-mass white dwarfs where oxygen is dredged-up \citep{Gaensicke2010}.  Also, small amounts of C, Al and Si  can be supported in the atmospheres of some stars by radiative levitation \citep{Dupuis2010}.}.
With the discovery of 
 17 individual elements within the photosphere of the   white dwarf GD 362, it  has become clear that we can use these externally-polluted stars to make detailed comparisons of the bulk compositions of extrasolar planetesimals with
those of the solar system's terrestrial planets \citep{Zuckerman2007}.

 About 94\% of bulk Earth's  total mass is contained within four dominant elements: O, Mg, Si and Fe \citep{Allegre2001}.  
Ground-based optical spectra with some additional results from {\it FUSE}, the {\it Far Ultraviolet Spectroscopic Explorer}, indicate that these same four species comprise at least 85\% of the heavy element mass in six highly polluted white dwarfs \citep{Klein2010, Klein2011, Dufour2010, Vennes2010, Zuckerman2010, Farihi2011, Melis2011}.  However, these optical studies  incompletely sample
 important species which only have strong lines in the ultraviolet.  We have therefore acquired spectra with the {\it Hubble Space Telescope} 
 to measure the abundances of additional elements in GD 40 and G241-6, two of the three most
 heavily polluted DB white dwarfs within 80 pc of the Sun \citep{Jura2012}.  Our ultraviolet data are a substantial improvement over older,
 lower-resolution data for GD 40 \citep{Friedrich1999, Wolff2002}; no previous ultraviolet spectroscopy has been performed
 on G241-6.
 
 GD 40  shows excess infrared continuum emission from a circumstellar dust disk \citep{Jura2007} that also displays 10 ${\mu}m$ silicate emission \citep{Jura2009a}.    While G241-6 and GD 40 have nearly the same effective temperatures, gravities and abundances of heavy-element pollutants \citep{Zuckerman2010, Klein2011}, G241-6 
  does not display an  infrared excess \citep{Xu2012}.  Either G241-6's disk is fully  gaseous or the parent body has been largely accreted and we are witnessing
  lingering contamination in the stellar photosphere; heavy elements take longer than  2 ${\times}$ 10$^{5}$ yr to sink below the outer mixing zone \citep{Koester2009,Klein2010}.
  
 In the standard picture \citep{Wetherill1990,Armitage2010}, planet-forming disks have well defined radial temperature zones and
 rocky planets assemble from material in a relatively narrow annulus.  Therefore a single condensation temperature
can usefully describe the bulk abundances.   This picture has been
adopted in models for the formation of extrasolar rocky planets \citep{Bond2010}; extrasolar planetesimals
are expected to be volatile-poor if they form sufficiently close to their host star.  

In the solar system, an element's condensation temperature is key in determining its incorporation into a planetesimal.  Typically, for the pressure, temperature and presumed composition of the solar nebula, the molecular constituents are computed and the fractional condensation is determined for each compound.   For bulk Earth, refractory elements
that are incorporated into  solids at temperatures above 1100 K are undepleted relative to solar abundances while  volatile elements with a condensation temperature less than 1000 K are  deficient  by factors of as much as 100 or more compared to a solar composition \citep{Allegre2001}.  The most primitive meteorites, the CI chondrites, only display depletions relative to solar values of elements with condensation temperatures below 200 K, while the more evolved CV chondrites have  depletions  by at least a factor of  two of elements with a condensation temperature below ${\sim}$700 K \citep{Wasson1988}.   

Previous support for  volatile-depletion within extrasolar planetesimals is given by the evidence that carbon is usually deficient  \citep{Jura2006, Farihi2009}.  Furthermore, \citet{Jura2012} found that, in aggregate, planetesimals
accreted onto DB white dwarfs have a factor of 10 less H$_{2}$O by mass than do CI chondrites, 
consistent with a scenario where these particular extrasolar rocky bodies were assembled interior to a   snow line which, depending upon the gas pressure, is located where the temperature is 
between 145 K and 170 K \citep{Lecar2006}.      By measuring volatile compositions of
individual extrasolar asteroids, we can more exactly assess the degree to which they are thermally processed.

The most important volatile elements that  might  be detected in polluted white dwarf spectra are C, N, and S  with 50\% condensation temperatures of 40 K, 123 K, and 664 K, respectively \citep{Lodders2003} because they are expected to be largely contained
within CH$_{4}$, NH$_{3}$ and FeS.
Another abundant volatile is oxygen, but this element  is special because much of it is incorporated into oxides of refractory elements such as Si and Mg.  Typically, there is ``leftover" oxygen which does not condense into solids
until the system attains a much lower temperature at which point H$_{2}$O ice forms.

In Section 2, we report  our data acquisition and reduction.  In Section 3, we report  atmospheric abundances while in Section 4 we infer compositions of the accreted parent
bodies.    In Section 5, we discuss our results in the context
of models for the growth of planetesimals and in Section 6 we present our conclusions. 

\section{DATA ACQUISITION AND REDUCTION}

For both GD 40 and G241-6, we obtained spectra with the G130M grating of the {\it Cosmic Origins Spectrograph} (COS) during Cycle 18 of the {\it Hubble Space Telescope} under GO Program 12290. Observations were obtained in the TIME-TAG mode using the 2{\farcs}5 diameter primary science aperture (PSA) with a central wavelength of 1300 {\AA}. Data were acquired during 3 spacecraft orbits for each star with a total integration time of 7,740 s for GD 40 and 8,835 s for G241-6. In each orbit, the FP-POS value was set to a different value to minimize the fixed pattern noise in the detector. The spectra extend from 1140 {\AA} to 1440 {\AA} except for a slight gap of $\sim$ 12 {\AA} near 1300 {\AA}; the spectral resolution was $\sim$ 20,000. 

The raw COS data were processed with CALCOS pipeline 2.15.4 and then smoothed with a boxcar of five pixels. At most wavelengths, the signal-to-noise ratio for each star is $>$ 25. The final spectra presented here are flux calibrated  and  in vacuum wavelengths. Because of strong terrestrial  day airglow emission from O I between 1300 {\AA} and 1308 {\AA} and from H I  near Lyman ${\alpha}$ \citep{Feldman2001}, we use the {\it timefilter} module to extract the night time portion of the data and rerun CALCOS to obtain the calibrated spectra for these two spectral intervals.  The airglow emission is completely removed at the oxygen lines; however,  residual emission remains
at Lyman ${\alpha}$ (Figure 1).   Near the edge of the spectrum at 1300 {\AA}, the pipeline reduction becomes somewhat noisy and unreliable (private communication, COS help desk).  The COS-measured flux of both stars near 1400 {\AA} is 1.4 ${\times}$ 10$^{-14}$ erg cm$^{-2}$ s$^{-1}$ {\AA}$^{-1}$,  the same as measured with  the GALEX satellite in a broad band centered at 1528 {\AA}.

Similar to the data reduction procedures described in \citet{Klein2010,Klein2011} and \citet{Zuckerman2010}, with IRAF, each absorption line was fit with a Voigt profile and the  equivalent width (EW) was derived. 
For each line, we performed three different measurements and the final EW uncertainty was obtained by adding three factors in quadrature: the standard deviation of the three fits, the average uncertainty of an individual fit to the assumed profile and the uncertainty of the continuum. For marginally blended lines, we used ``splot" in IRAF to perform deblending and measure the individual EW.
 We list in Table 1 the lines with values of the EW greater than 50 m{\AA} in at least one of
our two target stars.  

For elements without positive detections, we place upper limits to the EW by measuring the signal-to-noise ratio at the  region of its expected strongest  line. 
We calculate the 1 $\sigma$ uncertainty by dividing the  Full Width Zero Intensity by the signal-to-noise ratio.  We quote 3${\sigma}$ values for the upper limits to the EWs.   In the best-case,  this procedure translates to an EW of 24 m{\AA}, but, for some lines, the upper limit to the EW is 30 m{\AA} or even 35 m{\AA}.    We validated this procedure by inserting
artificial lines into the data.  

As expected from ground-based optical spectra, the ultraviolet spectra of  GD 40 and G241-6 each exhibit hundreds of photospheric absorption lines, many more than can be found in the older ultraviolet data of GD 40 which had only 1 {\AA} resolution \citep{Friedrich1999} and a lower signal-to-noise ratio.  The EW's of almost all the photospheric
lines are less than 200 m{\AA} and therefore would have been difficult to detect in the older data.

The average heliocentric velocity of the photospheric lines from the COS spectra is 19 km s$^{-1}$ and -23 km s$^{-1}$ for GD 40 and G241-6, respectively. As with the optical data, the dispersion  is near 5 km s$^{-1}$ for each star \citep{Klein2010, Klein2011}.  These values are different from the optically-measured values; \citet{Klein2010} and \citet{Zuckerman2010}  reported the velocities of 10 km/s and -28 km/s for GD 40 and G241-6, respectively. The difference between the optical and ultraviolet results can be attributed to the well-known difficulty of measuring absolute velocities with COS to better than
15 km s$^{-1}$ (COS Instrument Handbook).  In 
Figures 1-10, some of the observed spectra are additionally  shifted beyond the nominal correction for the heliocentric velocity by as much as  ${\pm}$ 4 km s$^{-1}$ to match more exactly the model spectra.  

\section{ABUNDANCE DETERMINATIONS}
Our abundance determinations from the ultraviolet data follow closely the procedures described previously for our optical observations \citep{Klein2010, Klein2011, Zuckerman2010}.  Unless otherwise specified, we use atomic data taken from either  VALD\footnote{    http://vald.astro.univie.ac.at/~vald/php/vald.php?docpage=usage.html}
or NIST\footnote{    http://physics.nist.gov/PhysRefData/ASD/Html/verhist.shtml} and the model atmosphere parameters listed in Table 2.
 For GD 40, \citet{Bergeron2011} derive an effective temperature of 14,780 K and $\log$ $g$ = 8.13 (cgs units).  Following the arguments in \citet{Klein2010} we adopt a temperature of 15,300 K which, however, may be in error by ${\sim}$ 500 K.  
Regardless, as shown in \citet{Klein2010, Klein2011}, variations of the effective temperature of 500 K and $\log$ $g$ of 0.3 typically lead to changes in the 
abundances of heavy elements relative to each other by less than 0.1 dex although the  abundances relative to He, the dominant gas in the photosphere, can vary by as much as 0.4 dex.  Therefore, the uncertainties in our adopted 
effective temperature and gravity do not lead to any qualitative change in our results.  For G241-6, our value of the effective temperature of 15,300 K is nearly equal to the 15,230 K temperature independently derived by \citet{Bergeron2011}.  For this same star, our value of
the gravity of $\log$ $g$ = 8.0 is slightly smaller than the value of $\log$ $g$ of \cite{Bergeron2011}. 

To derive abundances, we compare COS data with a model atmosphere to find the best fit to each line profile and EW.   
 For all detailed comparisons displayed below, we convolve the model predictions with
the COS line spread function \citep{Kriss2011}.  From the measured EW of each individual line, we compute
an  abundance estimate.  For each element, we then average, weighted by the quality of the data and atomic  parameters,  over each individual value to derive the ultraviolet-determined abundance that best fits all of an element's spectral lines.

The error determinations arise from scatter in the measurement of the equivalent widths and uncertainties in the models and atomic parameters.  
As demonstrated in \citet{Klein2010}, the abundance uncertainties propagated from EW determination are much less than the standard deviation of the abundance determined from different lines.  As a result, the  final abundance uncertainty  for each element  is derived from the internal dispersion of all its lines.  Because of modeling uncertainties, we assume a minimum error of 0.10 dex in all of our ``final"
abundances -- including those with only optical determinations.  

  Our final abundance for each element is determined by combining our previous optical studies \citep{Klein2010,Zuckerman2010} with our current COS analysis; these  results are
provided in Tables 3 and 4.  Below, we discuss each element in detail.  
Representative data are shown in Figures 1-10 where we also display the predictions from models employing the final set of abundances. 
\subsection{Potential Interstellar Lines}

Ultraviolet spectra of nearby stars display many interstellar absorption lines from species in the ground electronic state \citep{Spitzer1975}.  Using  {\it FUSE} to perform a survey of white dwarfs within 200 pc of the Sun, \citet{Lehner2003}  found, as expected, that interstellar absorption lines are pervasive.  Therefore, although both GD 40 and G241-6 lie within the local interstellar  bubble where the density is relatively low, their ultraviolet spectra must possess interstellar absorption lines. Extensive optical
surveys of the local interstellar gas show that it is patchy \citep{Welsh2010}, and we cannot predict its amount.  We therefore estimate the importance
of interstellar lines by considering a likely range of absorption line strengths.  A convenient tabulation of the equivalents widths of interstellar
absorption lines has been presented for ${\alpha}$ Vir \citep{York1979}, a star whose distance of 77 pc suggests
that it should experience roughly comparable amounts of interstellar absorption as GD 40 and G241-6.  The interstellar lines in our target  white dwarfs almost
certainly are weaker than those measured in ${\zeta}$ Oph \citep{Morton1975}, a star with famously strong interstellar lines whose distance of 112 pc puts it outside the local bubble.  In Table 1, in addition to listing the strengths of the photospheric lines of GD 40 and G241-6 from COS data, we list the measurements  from the {\it Copernicus} satellite of EWs in the spectra of ${\alpha}$ Vir and ${\zeta}$ Oph whose narrow lines
must be interstellar because the stars are rapidly rotating.

Another way to independently determine whether a line is interstellar or photospheric is to measure its radial velocity, since  these two types of lines usually have different velocities.   Unfortunately, for GD 40 these two velocities are almost the same so this method fails to discern interstellar lines.  The radial velocity offset between interstellar and photospheric lines for G241-6 is about 20 km s$^{-1}$ and can be used to identify interstellar features.  The  offsets for C and N are provided  in the  captions to Figures 2 and 3.

\subsection{Hydrogen}

Even though as shown in Figure 1 there is substantial  residual airglow emission, we have used the COS ``night-time" observations to derive upper bounds to the column density of atomic hydrogen, N(H), from the Lyman ${\alpha}$ line, H I 1216 {\AA}.   Unfortuantely, we cannot easily determine how much of
the line is interstellar and how much is photospheric.  For GD 40, the star with the stronger Lyman ${\alpha}$ line, if the hydrogen is entirely photospheric, then in the usual notation where 
[X]/He] denotes $\log$ $n$(X)/$n$(He), we derive [H]/[He] = -5.1,
approximately a factor of 10 greater than found from  H${\alpha}$  \citep{Klein2010}.    If [H]/[He] = - 6.14 as determined from the optical data, then following
 \citet{Bohlin1975}, we find N(H) = 1.0 ${\times}$ 10$^{20}$ cm$^{-2}$, approximately a factor of 10 greater than values of N(H) for
stars 70 pc from the Sun  \citep{Lehner2003}.  Therefore, it is likely that the Lyman ${\alpha}$ line is primarily photospheric. It is possible that the broad, weak H${\alpha}$ line was not well measured in the Keck HIRES echelle
spectrum which is difficult to flatten \citep{Klein2010}.  However, \citet{Voss2007} also report a very weak H${\alpha}$ absorption line and derived [H]/[He] = -6.02.  \citet{Bergeron2011} report [H]/[He] = -6.1, but with an uncertainty of 1.0 dex because the line is weak and difficult to measure.   Our
final estimate of the abundance reported in Table 3 has a large uncertainty.

We hope to reconcile the discrepancy between the optical and ultraviolet results for  photospheric hydrogen in GD 40 in future studies;  for now it is  mystery.  Fortunately this factor of 10 uncertainty in the hydrogen abundance does not materially affect the overall structure of the model atmosphere, so the  abundance determinations for other elements are unaffected.

For G241-6, from the lack of H${\alpha}$ absorption, \citet{Zuckerman2010} report  [H]/[He] $<$ -6.1 while \citet{Bergeron2011} report an upper limit to [H]/[He] of -5.43.  If the Lyman ${\alpha}$ line is entirely photospheric in origin, then [H]/[He] = -5.6, in disagreement with the upper limit
from \citet{Zuckerman2010}.  If, however,  [H]/[He] = -6.1 in the photosphere, then, as illustrated in Figure 1, N(H) = 3 ${\times}$ 10$^{19}$ cm$^{-2}$. Given that interstellar nitrogen lines detected toward this star are stronger than those measured for ${\alpha}$ Vir, it is plausible that the interstellar hydrogen column density could approach this value of 3 ${\times}$ 10$^{19}$ cm$^{-2}$.  If so, then  there is no  disagreement between the  measurement of [H]/[He] derived from the ultraviolet and the upper bound derived from the optical.

\subsection{Carbon}
The ultraviolet  lines of C I and C II that are accessible in our COS data  are subject to interstellar contamination because they arise either from the ground state or a very low-lying energy level.   In Figure 2, we display the comparison between our data and models
 for GD 40 and G241-6 for the spectral interval 1327 {\AA} to 1338 {\AA}.  In both stars, we see C II absorption at 1334.53 {\AA} and 1335.71 {\AA}.  However, neither interstellar nor photospheric absorption from C I is detected.  
 
In Table 1, the C II  lines toward GD 40 are somewhat stronger than even in ${\zeta}$ Oph.  Furthermore, the line at 1335.71 {\AA}
 arises from an excited fine structure level (see column 2 in Table 1) which is fully populated according to its statistical weight in a stellar photosphere but not so in the interstellar medium where the density
 is vastly lower.  Consequently,  an interstellar line at 1335.71 {\AA} is almost always weaker than the line at 1334.53 {\AA}.  In view
 of this consideration, it is likely that the C II lines toward GD 40 are largely photospheric with an abundance of [C]/[He] = -7.8. Our upper bounds to the CI lines near 1277 {\AA} and 1328 {\AA} are consistent with this result.
  
 For G241-6, it appears that the CII line at 1334.53 {\AA} is largely interstellar because it is offset
 from the photospheric velocity and because it is substantially stronger than the CII 1335.71 {\AA} line.    If   C II 1335.71 {\AA} is entirely photospheric, then [C]/[He] = -8.5.  Because the line is likely at least partly interstellar, we  conclude that this
 value of the abundance is  only  an upper limit.  Our upper bounds to the CI lines near 1277 {\AA} and 1328 {\AA} are consistent with this result.
 
\subsection{Nitrogen}
The accessible photospheric lines of N I at 1199.55 {\AA}, 1200.22 {\AA} and 1200.71 {\AA} all arise from the ground state and are strong in the interstellar medium.  A further difficulty is that these lines can be blended with photospheric Fe II 1199.67 {\AA},1200.24 {\AA} and 1200.75 {\AA}. As shown  in Figure 3, the Fe lines dominate this spectral region, and  there is no evidence for any nitrogen lines in the spectrum of GD 40; we derive [N]/[He] $<$ -8.8.  For G241-6, the observed lines probably arise from interstellar
absorption because  their wavelengths are offset and from what is expected in the stellar rest frame and, instead,  agree with the offset of the likely interstellar line at C II 1334.53 {\AA}.  There is no evidence for photospheric nitrogen; we derive an upper limit 
of  [N]/[He] $<$ -8.9.
\subsection{Oxygen}

We detect O I 1152.15 {\AA} and 1217.65 {\AA} in both stars; the data are shown in Figures 4 and 1.  These lines must be photospheric because they arises from  excited electronic states. Using night-time data only to avoid
day glow emission, we also detect O I 1302.17 {\AA}, 1304.86 {\AA} and 1306.03 {\AA}; these lines are shown in Figure 5.   Because the O I 1302.17 {\AA} line lies
near the edge of the array, the data are especially noisy and uncertain in its blue wing (private communication, COS help desk). The two longer wavelength O I lines are blended with lines of Si II.   Because of
these problems, we do not employ any of the three lines near 1302 {\AA} in our abundance determination
 \subsection{Aluminum}
 The only line that we have used for the abundance determination for this element is Al II 1191.80 {\AA} which cannot
 be interstellar because it arises from an excited energy level. The  fits to the data are shown
 in Figure 6.      As shown in Figure 1, we have  apparently detected
 Al II 1211.90 {\AA} in GD 40, but this line is blended with a feature from Fe II, we do not attempt to use these data to infer the aluminum abundance.  
 
 \subsection{Silicon}
We have detected  numerous ultraviolet lines of Si II and one line of Si III in both GD 40 and G241-6;  results are shown in Figures 5-8.  For GD 40, the Si II data cannot be consistently fit.  The model with our final abundance still under-predicts the strengths of the lines near 1260 {\AA} and 1350 {\AA}, but as reported in \citet{Klein2010} would over-predict the optical EWs.  The lines near 1194 {\AA} are well fit with the compromise ``final" value of [Si]/[He] given in Table 3 which is larger
than previously reported by \citet{Klein2010} by 0.32 dex.  In contrast, we find a more self-consistent fit of all the Si lines for G241-6 with only a 0.16 dex increase in [Si]/[He].    Independently, in GD 61, a comparably externally-polluted DB white dwarf, the optical and ultraviolet determinations agree to within 0.15 dex \citep{Farihi2011,Desharnais2008}.   \citet{Bautista2009} have re-examined the oscillator strengths of Si II, and while their values for the ultraviolet lines are close
to those employed in this paper, their recommended  values for the optical lines are about 0.3 dex weaker than we previously used and thus accounting for much of the discrepancy between the optical and ultraviolet abundance determinations for GD 40.   The abundance difference between Si and Mg
discussed in detail by \citet{Klein2010} for GD 40 is still present with our current analysis but is significantly less pronounced than previously reported.  

\subsection{Phosphorus}

Unequivocal detections  of three P II lines  for both stars are shown in Figure 4.    The stellar abundance determinations in Tables 3 and 4 are secure.  
\subsection{Sulfur}

As shown in Figure 7, we have certainly detected the S II doublet of 1204.27 {\AA} and 1204.32 {\AA}  for  G241-6 and perhaps for GD 40. These lines must be photospheric. For G241-6, we also detected the lines  at 1253.81 {\AA} and 1259.52 {\AA} (Figure 8); for GD 40, detections of the these two lines is marginal.   We also have detected S I 1433.28 toward G241-6 which  is unlikely to be interstellar.   The photospheric abundance determination of sulfur in G241-6 is secure; it is tentative in GD 40.  
\subsection{Chlorine}
 As shown in Figure 9, we only report upper limits to Cl I 1347.24 {\AA}.  
 
\subsection{Magnesium, Calcium, Titanium, Chromium, Manganese, Iron}

 We  detected lines of Mg, Ca, Ti, Cr, Mn  and  Fe in the COS spectra.  Our measurements for  Ti are not presented in Table 1 because they are all weaker than 50 m{\AA}.  For all of these elements, our final abundance determination largely is based upon our previous results. Except for Ca, the
abundances determinations  agree within the errors with those determined from the optical \citep{Klein2010,Zuckerman2010}.    Examples of the model fits for Fe and Mn are shown in Figures 2-4, 6-8 and 10.  For Ca, the only lines in the ultraviolet  have
poorly determined atomic parameters; they imply a significantly lower Ca abundance than derived from optical data.   Given the excellent fit to numerous optical lines, we do not reassess the
abundance of this element.

\subsection{Nickel}
We display detections of photospheric Ni in Figures 2 and 4.  The potentially interstellar Ni lines reported in Table 1 are so much stronger than observed in ${\zeta}$ Oph that
we assume that they originate in the photosphere and as-such, they are included in our abundance determination.

 \subsection{Copper, Germanium, Gallium}
 We display the spectra near the Cu II 1358.77 {\AA} line in Figure 10.  The line appears at the 3${\sigma}$ level in the spectrum of GD 40, and is at most marginally detected; we can only place an upper bound to
 the line in G241-6.   
  We have searched for  lines at  1404.12 {\AA} and 1237.07 {\AA}  for Ge and
 Ga, respectively.  They are undetected and not displayed here; the upper limits to the abundances are reported in Tables 3 and 4.

\section{PARENT BODY  COMPOSITIONS}

Having determined the atmospheric abundances, we now proceed to estimate the composition of the accreted parent bodies.
Because different elements gravitationally settle through the white dwarf's outer mixing zone at different rates, the relative abundances in the
photosphere may not be  the same as the relative abundances in the asteroidal parent body.  \citet{Koester2009} described
three phases of an accretion event. The first phase is a build-up of pollution after the initial creation of a circumstellar disk derived from
the tidal-disruption of an asteroid.   During this phase, which has a duration of approximately one settling time, the photospheric abundances are
the same as those in the parent body. For both GD 40 and G241-6, this phase has a duration of 2 ${\times}$ 10$^{5}$ yr.   During the second phase, there is an approximate steady state where the rate of
accretion from the disk is balanced by the loss at the bottom of the outer mixing zone. During this phase the concentration of each individual element in
the photosphere is determined by a balance between  its accretion rate and its settling time; derivation of parent body abundances must take this  effect into account.
The duration of this phase depends upon the mass of the parent body.  The estimated steady state accretion rates for GD 40
and G241-6 are 2 ${\times}$ 10$^{9}$ and 3 ${\times}$ 10$^{9}$ g s$^{-1}$ \citep{Klein2010, Zuckerman2010}.   Ceres, the most massive
asteroid in the solar system, has a mass of 9.1 ${\times}$ 10$^{23}$ g \citep{Baer2011}, and it is therefore imaginable that this phase could
have a duration of as much as 10$^{7}$ yr, but this is extremely uncertain.  As listed in Table 5, the inferred minimum mass of the asteroid accreted onto GD 40 of 3.8 ${\times}$ 10$^{22}$ g, slightly revised from the value of 3.6 ${\times}$ 10$^{22}$ g in \citet{Klein2010}, is much less than the mass of Ceres.   If the accretion from the disk varies with time \citep{Rafikov2011}, it may be better described to be in a quasi-steady state instead of an exact steady state.  In this case, the estimated mass fractions may
be uncertain by as much as a factor of two \citep{Jura2012}. After the disk is fully accreted, there is a final decay phase where the
light elements should dominate the photospheric composition since the heavy elements sink more quickly.  The duration of this final
phase is on the order of a few settling times, or, for GD 40 and G241-6,  ${\sim}$ 5 ${\times}$ 10$^{5}$ yr.

Because we are largely focused upon the upper limit to the  abundances of the volatile elements and because carbon and nitrogen  have particularly long settling times, it is most conservative  to assume that the systems are in the ``build-up" phase
because this is when we would infer the highest fractional abundances for the volatiles.  If, instead, the system is in a steady state or declining phase, we  overestimate the fraction of mass carried in the lighter elements and underestimate the fraction of mass carried in the heavier elements.   According to \citet{Klein2010}, the difference between derived
mass fractions between the build-up phase and the steady state phase is at most a factor of 1.6, not enough to qualitatively change our
conclusions.  If a system is in the declining phase, then the correction to the derived mass fraction of a heavy element such as Ni for gravitational settling can, in principle, be larger
than a factor of 10 \citep{Koester2009}.

 We display in Figures 11-13  the relative abundances listed in  Table 5.     
Because we have a comprehensive suite of abundances in the atmospheres of GD 40 and G241-6, we follow standard analysis of
solar system objects and display the fraction of the total mass of the parent body carried by each element.  

We see in Figure 11 that the two elements with 50\% condensation temperatures less than 150 K, well inside the snow line \citep{Lecar2006}, are significantly depleted in the parent bodies accreted onto GD 40 and G241-6 relative
 to the ``adjusted sun" and CI chondrites.  
In contrast, we see in Figures 12-13 that the elements with condensation temperatures greater than 1000 K have mass
fractions roughly similar to bulk Earth and CI and CV chondrites.  Sulfur with an intermediate 50\% condensation temperature of 664 K is  heavily depleted in GD 40 but undepleted in G241-6.

\section{DISCUSSION}

In the usual models for their formation in a disk orbiting a central star, rocky planets grow within a radial annulus.  The temperature
within this annulus controls the material which is incorporated into solids and therefore, ultimately, the chemical
composition of a planet \citep{Armitage2010}.  In this picture, there is a snow line, and planets formed interior to this demarcation
have little ice.  As a first approximation, the abundances within the rocky bodies in the solar system can be explained
by this scenario.  For example, 
based solely on solar abundances, carbon should be one of the most abundant elements within a rocky planet.  However,  in a cosmochemical environment, it is highly volatile and strongly depleted both in the inner solar system \citep{Lee2010} and  in extrasolar asteroids  \citep{Jura2006}.  Similarly, water-ice is much more abundant in the outer solar system than in the inner solar system \citep{Jewitt2007}.

The  depletions of C and S in the parent body accreted onto GD 40 are similar to those of bulk Earth while the upper bound of N in the accreted material is consistent with bulk Earth's low fractional mass of this element.   Also, the abundances of the refractories are mostly similar to those
in bulk Earth except for an unexplained enhancement of Ca and possibly Ti.  Therefore, the formation pathway  of the parent body that orbited GD 40 may have been analogous  to the  formation of Earth.  Although C and N are substantially depleted in the parent body accreted onto G241-6, the solar abundance of S suggests that
the material condensed at a temperature less than 600 K.  Therefore, the formation route of the parent body that orbited this star may have been more similar to that
of the  CV chondrites.    The volatiles deficiencies in GD 40 and G241-6 seem to be part of a more general pattern of the parent bodies accreted onto DB white
dwarfs in aggregate likely having formed interior to a snow line \citep{Jura2012}.  

Although similar, the abundances of GD 40 and G241-6 are not identical.  As seen in Table 5, while the dominant lighter elements -- O and Mg -- have nearly the same
abundance, the elements as heavy or heavier than  Ca are all more abundant in GD 40 by as much as a factor of 2.5. A natural explanation of this difference is that accretion onto G241-6 has effectively stopped and
that we are now witnessing a system in a decaying phase where the heavy elements gravitationally settle before the lighter elements.  This scenario is consistent
with GD 40's possession of a dust disk and the absence of any circumstellar dust orbiting G241-6.  Therefore, our ``conservative" approach in assuming  that G241-6 is in the ``build-up" phase may overestimate the mass fractions of the relatively light volatiles, but it correspondingly may underestimate the mass fractions of the relatively heavy refractories.  Sulfur and silicon have nearly the same atomic weight and gravitationally settle at nearly the same speed.  As illustrated in the modeling of \citet{Jura2012}, the imputed mass
fraction carried by silicon -- intermediate in mass between oxygen and iron -- is approximately constant during an episode where the accretion rate varies by more than a factor of 10.  Consequently, the inferred sulfur mass fraction is insensitive to whether the system is a build-up or steady-state phase.  Even if G241-6 is in a final, declining phase, the likely corrections to the relative abundances of the heavy elements are approximately 
a factor of two; such adjustments would not lead to a qualitative change in our analysis.  

We now briefly assess whether during either star's high-luminosity Asymptotic Giant Branch evolution prior to becoming
a white dwarf, the composition of the asteroids was dramatically altered by sublimation and the consequent loss of volatile
elements.  According to  idealized calculations \citep{Jura2010}, less than half of any initial internal H$_{2}$O is likely to be lost  during such a phase by an asteroid
with a radius greater than 60 km.   The parent body accreted onto GD 40 had a radius
of at least 120 km \citep{Klein2010}, and therefore sublimation during the pre-white dwarf evolution of the star is unlikely to explain the deficiency of volatiles in the accreted asteroid.  Because GD 40 has an infrared excess, it is likely that a single, large asteroid is responsible for the observed pollution; otherwise the dust particles in the disk likely would
have been destroyed by mutual collisions \citep{Jura2008}.  For G241-6, a similarly large mass is required to explain the observed pollution and probably the pollution was delivered by one large parent body \citep{Klein2011}. It seems
plausible that the inferred deficiencies of volatiles were established when the planetesimals formed.  Another possibility is that ${\sim}$100 km
asteroids are differentiated because of substantial heating by radioactive elements \citep{Young2003}.  Because the source of $^{26}$Al -- the primary source of heating \citep{Ghosh1998}  -- in the early solar system is unknown \citep{Tatischeff2010}, it is uncertain how to compute the thermal evolution in extrasolar planetary systems.  The degree to which complex, differentiated
asteroids with porosity and hydrated minerals can retain their water and other volatiles is unknown and needs to be addressed in future models.  

In the parent bodies accreted onto GD 40 and G241-6, a simple thermal processing model can  explain the bulk compositions.  However,  it is most  unlikely that all extrasolar asteroids are volatile depleted; \citet{Koester2011} report that
a suite of older DZ white dwarfs have, on average,  nearly solar sodium abundance.  This element has a 50\% condensation temperature of 958 K \citep{Lodders2003} and is deficient relative to solar in bulk Earth \citep{Allegre2001}.   Possibly these stars with cooling ages near 2 Gyr
are polluted by a different class of planetesimal than GD 40 and G241-6 which have cooling ages near 0.2 Gyr \citep{Bergeron2011}.   Elaboration of dynamic models for the orbital perturbations
of planetesimals such as those computed by \citet{Bonsor2011} or \citet{Debes2012} may allow us to determine when to expect
accretion of volatile-rich or volatile-poor planetesimals.  

Besides thermal processing, other processes such as differentiation may also be important \citep{Klein2010, Melis2011, Zuckerman2011} in determining
the composition of extrasolar asteroids.  It is most valuable in testing models for planetesimal formation and evolution to measure mass fractions of as many different
elements as possible.  As summarized by \citet{Zuckerman2011}, seven white dwarfs have been found to be polluted by  at least eight different
elements.  
 Future studies of such extensively polluted white dwarfs may elucidate different evolutionary pathways taken
by extrasolar planetesimals.

\section{CONCLUSIONS}

We have found  in the white dwarfs GD 40 and G241-6 that two highly  volatile elements -- carbon and  nitrogen -- 
comprise only a small fraction   of the mass  in the accreted parent bodies.  The deficiency of these two volatiles parallels
what is measured for bulk Earth and  previous studies showing that in aggregate, there was relatively little  water-ice embedded in the ensemble of planetesimals accreted onto nearby
externally-polluted white dwarfs with helium-dominated atmospheres   Sulfur is significantly depleted in GD 40 but not  G241-6.   In contrast, elements with condensation temperatures greater than 1000 K have  mass fractions similar to those measured in bulk Earth and chondrites.  This contrast between volatiles and refractories is consistent with scenarios where the composition of a planetesimal is strongly
dependent upon the radial location in the disk where it was formed.

This work has been partly supported both by NASA (for the ${\it HST}$ telescope time) and by  NSF and NASA grants to UCLA to study
polluted white dwarfs.

\newpage

 \begin{center}
 Table 1  -- Measured Equivalent Widths
 \\
 \begin{tabular}{lrrrrlllllll}
 \hline
 Line &  ${\chi}$ & GD 40 &  G241-6 &  ${\alpha}$ Vir$^{a}$ &  ${\zeta}$ Oph$^{b}$ \\
& (eV)  & W$_{\lambda}$(m{\AA})  & W$_{\lambda}$(m{\AA}) &W$_{\lambda}$ (m{\AA}) & W$_{\lambda}$(m{\AA}) & Comment \\
 \hline
 H I 1215.67 & 0.000 & 5700$^{c}$ (570)  & 4000$^{c}$ (400)    & $<$2130 &   16700   \\ 	  
 \\
 C II 1334.53 & 0.000 & 201 (10) &214 (11) & 112.1 & 189 \\
 C II 1335.71 & 0.008 & 212 (11) &127 (10)& 62.6 & 140 \\
 \\
 N I 1199.55 &  0.000& &113 (13) & 69.8  & 146 \\
 N I 1200.22 &0.000 &  &  134 (12)&  67.1 & 131 \\
 N I 1200.71 & 0.000 & & 147 (12)& 56.2 & 133 \\
 \\
 O I 1152.15 & 1.967 & 540 (19) & 448 (18) \\ 
 O I 1217.65 & 4.190 & 51 (25) & 76 (22) &  &&blend \\
 O I 1302.17 & 0.000 & 1130 (166)     &  966 (63)  &   89 & 201 & edge \\
 O I 1304.86 & 0.020 & 761 (169) & 508 (56) &  $<$ 0.7 & $<$ 3.9   & blend \\
 O I 1306.03 & 0.028 & 578 (49) & 423 (46) & $<$ 0.7 & $<$ 4.0 & blend \\
 \\
 Mg II 1239.93 & 0.000 & 151 (11)& 146 (12) & 1.0  & 18.7\\
 Mg II 1240.94 & 0.000 & 112 (10)  & 94 (10) & 0.2  &  13.2 \\
 \\
 Al II 1191.81 & 4.659 & 68 (9) & $<$ 60 & \\
 \\
 Si II 1190.42 & 0.000 & 1322 (28) & 846 (29) & 66.1 & 132 \\
 Si II 1193.29 & 0.000 & 1182 (99) & 1127 (158) &  & 147 \\
 Si II 1194.50 & 0.036 & 2587 (88) & 1744 (104) & 1.0 & 14.3 \\
 Si II 1197.39 & 0.036 & 1180 (37) & 610 (18) & & 4.8 \\
 Si II 1223.90 & 5.323 & 132 (17) & 118 (18) &&& u \\
 Si II 1224.25 & 5.323 & 132 (18) & 131 (13)   & &&u, blend \\
 Si II 1224.97 & 5.323 & 53 (21) & 52 (15) & &&u \\
 Si II 1226.80 & 0.098 & 199 (61) & 189 (31) &&& u,  blend \\
 Si II 1226.98 & 0.098 & 178 (40) & 192 (34) & &&u, blend \\
 Si II 1227.60 & 5.345 & 292 (21) & 245 (14) & &&u \\
 Si II 1228.74 & 5.323 & 422 (29) & 499 (24) & &&u, blend \\
 Si II 1229.38 & 5.345 & 425 (26) & 380 (18) &&& u \\
 Si II 1246.74 & 5.310 & 141 (18)  & 115 (12) \\
 Si II 1248.43 & 5.323 & 193 (13) & 166 (10) \\
 Si II 1250.09 & 6.858 & 125 (21) & 134 (26) & &&u,  blend \\
  Si II 1250.44 & 6.859 & 226 (18) & 296 (63) &&& u \\
 \hline
 
 \end{tabular}
 \end{center}
 \newpage
  \begin{center}
 Table 1  -- Continued
 \\
 \begin{tabular}{lrrrrlllllll}
 \hline
 Line &  ${\chi}$ & GD 40 &  G241-6 &  ${\alpha}$ Vir$^{a}$ &  ${\zeta}$ Oph$^{b}$ \\
& (eV)  & W$_{\lambda}$(m{\AA})  & W$_{\lambda}$(m{\AA}) &W$_{\lambda}$ (m{\AA}) & W$_{\lambda}$(m{\AA}) & Comment\\
 \hline
 Si II 1251.16 & 5.345 & 244 (15) & 176 (15) &&& u \\
 Si II 1260.42 & 0.000 & 2765 (85) & 2178 (96) & 88.5 & 170 \\
 Si II 1264.74 & 0.036 & 5542 (270) & 4676 (77) & 0.3 & 15.4 \\
 Si II 1304.37 & 0.000 &  614 (49)       & 500 (46)    & 59.9 & 132  & blend \\
 Si II 1309.28 & 0.036 & 1039 (45) & 813 (35) &   & $<$ 4 \\
 Si II 1346.88 & 5.323 & 157 (15) & 121 (15) \\
 Si II 1348.54 & 5.310 & 173 (13) & 133 (14) \\
 Si II 1350.07 & 5.345 & 284 (25) & 265 (14) \\
 Si II 1352.64 & 5.323 & 186 (22) & 144 (13) \\
 Si II 1353.72 & 5.345 & 162 (20) & 134 (13) \\
 \\
 Si III 1206.50 & 0.000 & 1163 (35) & 994 (33) & 58.7 & 123 & u \\
 \\
 P II 1154.00 & 0.058 & 82 (21) & 59 (14)\\
 P II 1159.09 & 0.058 & 17 (7) & $<$ 30 \\
 P II 1249.83 & 1.101 & 86 (18)  & 41 (9)\\
 \\
 S I 1425.03 & 0.000 & $<$ 37 & 78 (21) &  &\\
 \\
 S II 1204.28$^{d}$ & 1.845 & 36 (11) & 98 (13) &&& \\
 S II 1250.58&  0.000 & blend & 84 (18) & 20.1&  100 \\
 S II 1253.81 &0.000 & 37 (10) &159 (12) &  30.4 & 106 \\
 S II 1259.52 & 0.000 & $<$ 38 & 93 (20) &  49.2 & 112 & blend \\
 \\
 \\
 \\
 Ca II 1432.50 & 1.692 & 130 (32) & $<$ 38 \\
 Ca II 1433.75 & 1.700 & 90 (14) & 54 (9) \\
 \\
 \\
 Cr II  1434.98 & 1.483 & 57 (12) & \\
 Cr II  1435.20 & 1.506	& $<$ 36	&	\\
  \hline
  \end{tabular}
 \end{center}
\newpage
  \begin{center}
 Table 1  -- Continued
 \\
 \begin{tabular}{lrrrrlllllll}
 \hline
 Line &  ${\chi}$ & GD 40 &  G241-6 &  ${\alpha}$ Vir$^{a}$ &  ${\zeta}$ Oph$^{b}$ \\
& (eV)  & W$_{\lambda}$(m{\AA})  & W$_{\lambda}$(m{\AA}) &W$_{\lambda}$ (m{\AA}) & W$_{\lambda}$(m{\AA}) &Comment\\
 \hline
Fe II 1143.23 & 0.000 & 279 (37) & 97 (21) & $<$ 2 & 38 & edge\\ 
 Fe II 1144.94 & 0.000 & 240 (23) & 209 (24) \\  
Fe II 1146.83 & 0.107 & $<$ 50 & $<$ 50 &&& blend\\ 
 Fe II 1146.95 & 0.048 & 79 (26) & $<$ 50 &&& blend \\
Fe II 1147.41 & 0.083 & 275 (50) & 150 (34) & &&blend \\
Fe II 1148.08 & 0.107 & 182 (35) & 271 (90) &&& blend \\ 
Fe II 1148.28 & 0.048 & 173 (44) & 288 (27) &&& blend \\ 
Fe II 1149.59& 0.121 & 82 (22)  & 60 (21) \\
Fe II 1150.29 &	0.083 & 51 (16) 	&	$<$ 50 &&& blend\\ 
Fe II 1150.47 & 0.121 & 89 (18) & 88 (21) &&& blend \\ 
 Fe II 1150.69 & 0.083 & 67 (14) & $<$ 38 &&& blend \\ 
 Fe II 1151.15 & 0.083 & 127 (20) & 87 (14) \\
 Fe II 1154.40 & 0.121 & 125 (34) & 48 (13) &&& blend\\ 
 Fe II 1159.34 & 0.986 & 57 (17) & 48 (13)  &&& blend\\
 Fe II 1168.49 & 2.807 & 69 (18) & 49 (14) &&& blend \\ 
 Fe II 1169.19 & 0.986 & 63 (17) & $<$ 30& &&blend  \\ 
 Fe II 1175.68 & 0.232 & 99 (22) & $<$ 30 &&& blend \\ 
 Fe II 1183.83 & 0.986 & 68 (14) & 44 (11) \\
 Fe II 1187.42 & 0.986 & 62 (12) & 42 (11)&&& blend \\ 
  Fe II 1192.03 & 1.040 & 68 (25) & 50 (13)&&& blend \\
 Fe II 1198.93 & 2.635 & 65 (13) & 41 (12) \\
 Fe II 1267.42 & 0.048 & 124 (18) & 54 (14)& &&blend \\ 
 Fe II 1271.98 & 0.083 & 104 (16) & 54 (13) \\
 Fe II 1272.61 & 0.083 & 86 (13) & 44 (13) \\
 Fe II 1275.14 & 0.107 & 58 (9) & 32 (7) \\
 Fe II 1275.78 & 0.107 & 94 (10) & 80 (10) \\
 Fe II 1277.64 & 0.121 & 136 (12) & 93 (12) \\

    \hline
 
 \end{tabular}
 \end{center}

  \newpage
  \begin{center}
 Table 1  -- Continued
 \\
 \begin{tabular}{lrrrrlllllll}
 \hline
 Line &  ${\chi}$ & GD 40 &  G241-6 &  ${\alpha}$ Vir$^{a}$ &  ${\zeta}$ Oph${^b}$ \\
& (eV)  & W$_{\lambda}$(m{\AA})  & W$_{\lambda}$(m{\AA}) &W$_{\lambda}$ (m{\AA}) & W$_{\lambda}$(m{\AA}) & Comment \\
 \hline
   Fe II 1311.06 & 4.076 & 50 (11) & $<$ 40 \\
  Fe II 1358.94 &  3.245 & 44 (16) & $<$ 40 &&& blend \\
 Fe II 1361.37 & 1.671 & 77 (12) & 64 (9) \\
 Fe II 1364.58 & 3.267 & 128 (11) & 38 (10) &&& blend\\ 
 Fe II 1371.02 & 2.635 & 91 (11) & 68 (12) \\
 Fe II 1372.29 & 1.695 & 50 (10) & 40 (12) \\
 Fe II 1373.72 & 3.768 & 58 (10) & 51 (11) \\ 
Fe II 1375.17 & 2.657 & 80 (6) & 65 (12) \\
 Fe II 1377.99 & 3.814 & 46 (11) & $<$ 28 \\
 Fe II 1379.47 & 2.676 & 56 (12) & 64 (10) &&&blend \\ 
 Fe II 1379.62 & 1.724 & 44 (14) & 25 (9) & &&blend \\
 Fe II 1383.58 & 2.692 & 70 (7) & 46 (10) \\
 Fe II 1387.22 & 2.522 & 55 (11) & 30 (10) \\
 Fe II 1392.15 & 2.522 & 69 (10) & $<$ 30 \\
Fe II 1392.82 & 2.580 & 93 (13) & 50 (10) \\
 Fe II 1405.61 & 0.232 & 64 (14) & 64 (12)&&&blend  \\ 
 Fe II 1408.48 & 2.635 & 68 (15) & 42 (11) \\
 Fe II 1412.84 & 0.232 & 89 (16) & 49 (12) &&& blend \\ 
 Fe II 1413.70 & 2.657 & 62 (14) & $<$ 30 \\ 
 Fe II 1420.91 & 2.692 & 55 (14) & $<$ 40  \\
 Fe II 1424.72 & 0.301 & 79 (13) & 55 (16)& &&blend  \\
 Fe II 1435.00 & 0.352 & 58 (14) & 42 (13) & &&blend \\
  \\
 Ni II 1317.22 & 0.000 & 126 (11) & 88 (10)  &  \\
 Ni II 1335.20 & 0.187 & 108 (10) & 61 (9) \\
 Ni II 1370.13 & 0.000 & 110 (17) & 85 (15)& $<$ 6& 19.7\\
 Ni II 1381.29 & 0.187 & 79 (10) & 41 (10) \\ 
 Ni II 1393.32 & 0.000 &43 (10) & $<$ 30 & $<$9.1& $<$ 9.1  \\
 Ni II 1411.07 & 0.187 & 63 (12) & 33 (11) \\ 
 \hline
 \end{tabular}
 \end{center}
 In this Table, ${\chi}$ is the  lower energy of the observed transition, ``u" means that the atomic parameters are particularly uncertain while ``edge" means that the line lies very near the edge
 of the array.   The measurements for ${\alpha}$ Vir and ${\zeta}$ Oph  are for interstellar lines and are 
 provided for comparison with our COS-measured values given for GD 40 and G241-6.
 \\
$^{a}$from \citet{York1979}
 \\
 $^{b}$from \citet{Morton1975}
 \\
  $^{c}$Derived from the model fit shown in Figure 1 rather than measured directly from the data
 \\
 $^{d}$blend of S II 1204.27 {\AA} and S II 1204.32 {\AA}; atomic parameters from \citet{Tayal2010}
 \newpage
\begin{center}
Table 2 -- Stellar Parameters
\\
\begin{tabular}{lllllllll}
\hline
Property & GD 40 & G241-6 & Reference \\
\hline
$T_{*}$ (K) & 15,300 & 15,300 & (1), (2) \\
$\log$ $g$ (cm$^{2}$ s$^{-1}$)& 8.0 & 8.0 & (1), (2) \\
$D$ (pc) & 64 & 65 & (3)\\
\hline
\end{tabular}
\end{center}
(1) \citet{Klein2010}; (2) \citet{Zuckerman2010}; (3) \citet{Bergeron2011}

 \newpage
 \begin{center}
 Table 3 -- Photospheric Abundances: GD 40
 \\
 \begin{tabular}{lllllll}
 \hline
 $Z$ &  Optical & Ultraviolet & Final \\
    \hline
 H &  -6.16$^{+0.14}_{-0.20}$   &  -5.1$^{+0.2}_{-1.0}$& -5.1$^{+0.2}_{-1.0}$  \\
 \\
 C & ... & -7.8 ${\pm}$ 0.2 &  -7.8 ${\pm}$ 0.20  \\
 \\
 N & $<$-5.3& $<$-8.8 & $<$ -8.8 \\
 \\
 O &   -5.61$^{+0.08}_{-0.09}$ &  -5.68 ${\pm}$ 0.10 &    -5.62 ${\pm}$ 0.10&\\
 \\
 Mg   & -6.24$^{+0.12}_{-0.16}$ & -6.05 ${\pm}$ 0.10 &      -6.20 ${\pm}$ 0.16  \\
 \\
 Al &  $<$-7.1 &      -7.35 ${\pm}$ 0.12& -7.35 ${\pm}$ 0.12 \\
 \\
 Si &   -6.76$^{+0.07}_{-0.08}$     &    -6.28 ${\pm}$  0.10     &        -6.44 ${\pm}$ 0.30 &  \\
 \\
 P&$<$-8.2 &  -8.68 ${\pm}$ 0.13 & -8.68 ${\pm}$ 0.13 \\
 \\
 S  & ... &-7.8: ${\pm}$ 0.20 & -7.8: ${\pm}$ 0.20  \\
 \\
 Cl & ... & $<$-9.06 & $<$-9.06\\
 \\
 Ca & -6.88$^{+0.14}_{-0.22}$ & -7.37 (0.20) & -6.9${\pm}$0.2 \\
 \\
 Ti & -8.61 $^{+0.13}_{-0.20}$ &  ... &  -8.61 $^{+0.13}_{-0.20}$ \\
 \\
 Cr & -8.31$^{+0.12}_{-0.16}$& ... &    -8.31$^{+0.12}_{-0.16}$  \\
 \\
 Mn & -8.62$^{+0.13}_{-0.18}$ &    -8.73 ${\pm}$ 0.10 &      -8.64 ${\pm}$0.14 \\
 \\
 Fe &  -6.48$^{+0.09}_{-0.12}$ & -6.46 ${\pm}$ 0.10 &  -6.47 ${\pm}$ 0.12   \\
 \\
 Ni  & ${\leq}$-7.7  &-7.84 ${\pm}$ 0.26 &  -7.84 ${\pm}$ 0.26    \\
 \\
 Cu & ... & $<$-9.91 & $<$-9.91 \\
 \\
  Ga & ... &$<$-10.8  & $<$-10.8 \\
  \\
 Ge & ... & $<$-10.2 &$<$-10.2 \\
  \hline
  \end{tabular}
 \end{center}
 
 Abundances are given  ${\log}$[$n$(Z)/$n$(He)].
\newpage

 \begin{center}
 Table 4 -- Photospheric Abundances: G241-6
 \\
 \begin{tabular}{lllllll}
 \hline
 $Z$ &  Optical & Ultraviolet & Final \\

    \hline
 H & $<$-6.1 &  -5.9 ${\pm}$ 0.3& -5.9 ${\pm}$ 0.3 & \\
 C &...  & $<$-8.5&  $<$-8.5\\
 N & ... &  $<$-8.9& $<$-8.9\\
 O & -5.6 ${\pm}$ 0.10&  &-5.64 ${\pm}$ 0.11  \\
 Mg & -6.29 ${\pm}$ 0.05 &  -6.14 ${\pm}$ 0.10      &   -6.26 ${\pm}$ 0.10  \\
 Al &  $<$-7.0 &$<$-7.7 &$<$-7.7 \\
 Si &-6.78 ${\pm}$ 0.06 &  -6.55 ${\pm}$ -0.10   & -6.62 ${\pm}$ 0.20\\
 P& ... &  -9.04 ${\pm}$ 0.13   & -9.04 ${\pm}$ 0.13  \\
 S &... & -7.07 ${\pm}$ 0.28 &-7.07 ${\pm}$ 0.28 \\
 Cl & ... &  $<$-9.2 & $<$-9.2 \\
 Ca   &-7.25 ${\pm}$ 0.07& -7.9 (0.2) & -7.3 ${\pm}$ 0.20 \\
 Ti &  -8.97 ${\pm}$ 0.04& ... & -8.97 ${\pm}$ 0.10 \\
 Cr & -8.46 ${\pm}$ 0.05& ... & -8.46 ${\pm}$ 0.10 \\
 Mn & -8.8 ${\pm}$ 0.10 &  -8.74 ${\pm}$ 0.10   & -8.76 ${\pm}$ 0.20\\
 Fe & -6.76 ${\pm}$ 0.06 & -6.86 ${\pm}$ 0.10   &  -6.82 ${\pm}$ 0.14  \\
 Ni &$<$ -7.5 & -8.15 ${\pm}$ 0.40&-8.15 ${\pm}$ 0.40\\
 Cu & ... & $<$-9.90 & $<$-9.90& \\
  Ga &... & $<$-10.8  & $<$-10.8 \\
 Ge &... &  $<$-10.2  &$<$-10.2\\
  \hline
 \end{tabular}
 \end{center}
 Abundances are given as  ${\log}$[$n$(Z)/$n$(He)].
  \newpage
 \begin{center}
 Table 5 --  Mixing-Zone Masses 
 \\
 \begin{tabular}{lll}
 \hline
 $Z$ &   GD 40   &G241-6  \\
    &   (10$^{21}$ g) &  (10$^{21}$ g)\\
    \hline
 H & 3.4 &   0.54      \\
 C & 0.081  & $<$0.016  \\
 O &  16 & 16   \\
 Mg & 6.5 &   5.7   \\
 Al &  0.51 &$<$ 0.23  \\
 Si  & 4.3 & 2.9  \\
 P&0.027 &  0.012   \\
 S &   0.22: &  1.2 \\
 Ca & 2.1 &  0.85 \\
 Ti &  0.050 & 0.022\\
 Cr &   0.11 &0.077 \\
 Mn &   0.055 & 0.041 \\
 Fe &   8.0 &3.6  \\
 Ni & 0.36&0.18 \\
  \hline
  Total &       38    &  31 \\
 \end{tabular}
 \end{center}
   We assume the total mass of
 the outer convection zone is 1.7 ${\times}$ 10$^{27}$ g \citep{Klein2010}.  The value for the ``Total" mass of pollutants in the mixing zone excludes hydrogen because some of this element may
 be primordial  \citep{Bergeron2011} or it may have been accreted in previous episodes.  The mass fraction of each individual element is derived   by computing the ratio of its mass to that of the ``Total"; the results are displayed in Figures 11-13.
 The correspondence  between the mass fractions of pollutants in the stellar atmosphere and the bulk composition of the accreted parent body is discussed in Section 4.

 \begin{figure}
 \plotone{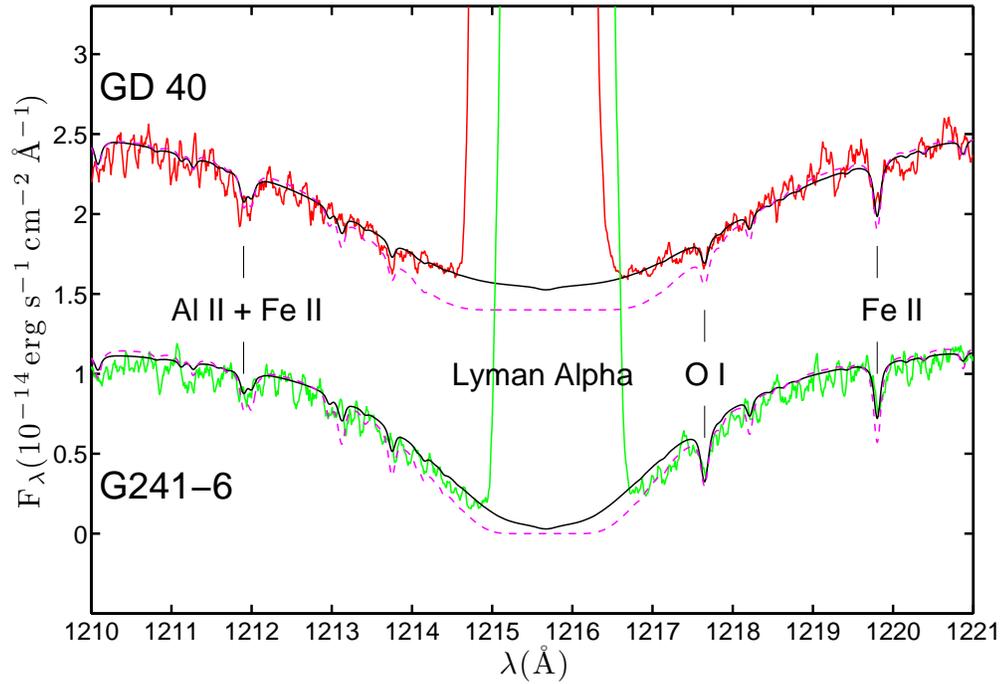}
 \caption{Lyman ${\alpha}$ absorption in GD 40 and G241-6.  The black lines represent models with the ``final" abundances given in Table 3 and 4 without any interstellar contribution. The dashed pink lines represent models with 
 [H]/[He] = -6.16 and column densities of interstellar hydrogen of  1 ${\times}$ 10$^{20}$ cm$^{-2}$ and 3 ${\times}$ 10$^{19}$ cm$^{-2}$ toward GD 40 and G241-6, respectively (see Section 3.2).  For clarity, the spectrum
 for GD 40 is offset by 1.5 ${\times}$10$^{-14}$ erg s$^{-1}$ cm$^{-2}$ {\AA}$^{-1}$.  Because the data are shifted into the star's frame of reference, the dayglow
 emission lines for the two stars are displaced.  }
 \end{figure}

 \begin{figure}
  \plotone{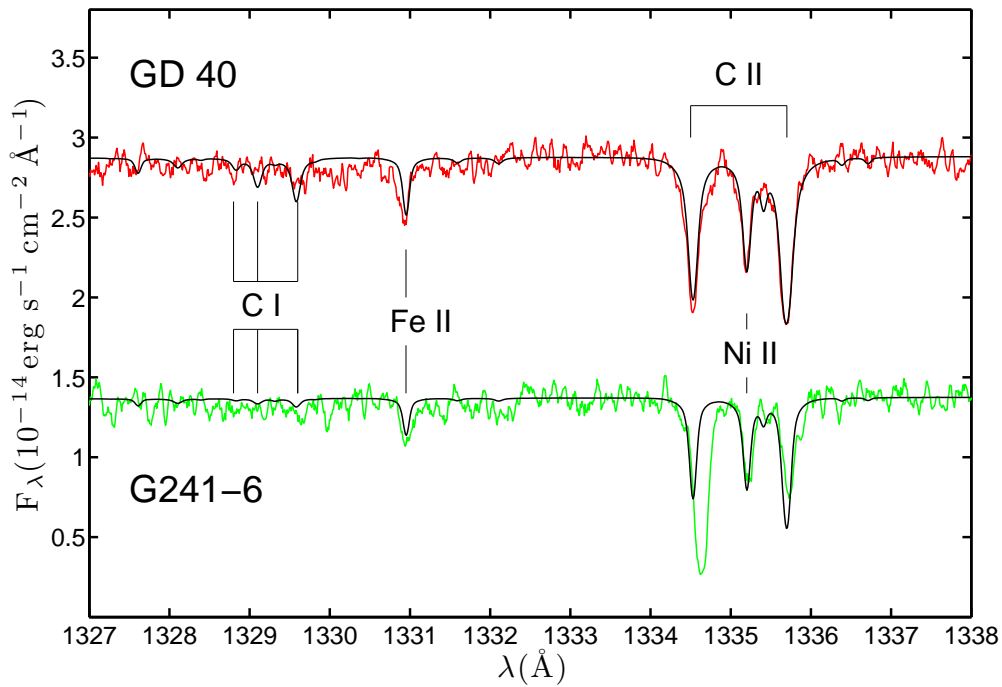}
 \caption{Similar to Figure 1 except for C I and C II. For G241-6, the C II line at 1334.53 {\AA} is shifted by +19 km s$^{-1}$ with respect to the photospheric velocity; likely it is largely interstellar. We regard the C I lines to be undetected in both stars.}
 \end{figure}

  \begin{figure}
  \plotone{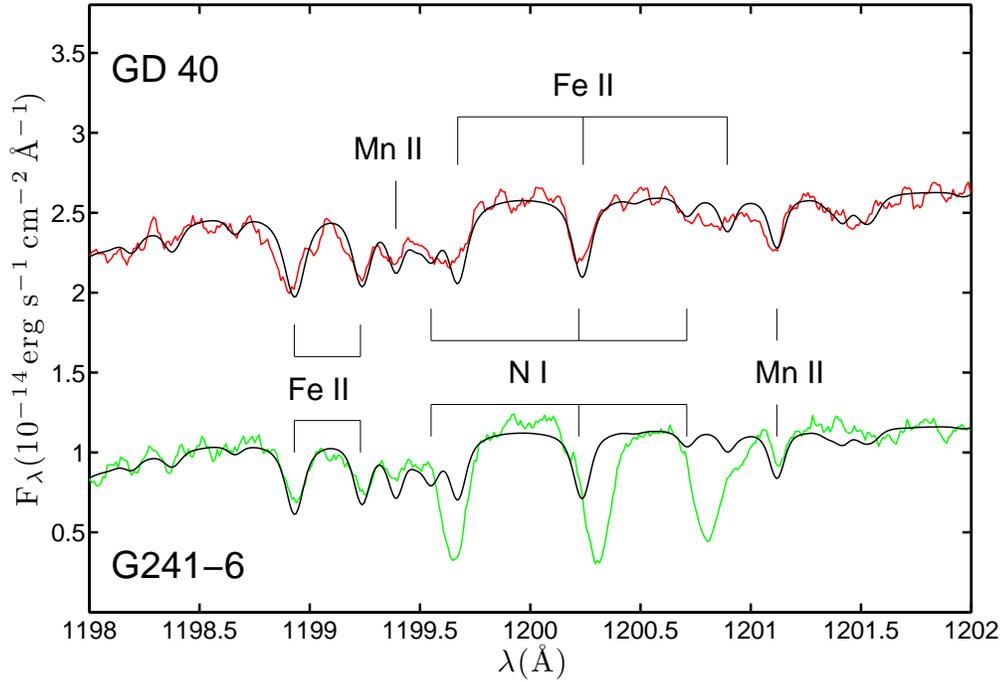}
 \caption{Similar to Figure 1 except for  the spectral region near prominent N I lines.  For G241-6, the N I lines at 1199.55 {\AA}, 1220.22 {\AA} and 1200.71 {\AA} are shifted by +26 km s$^{-1}$, +21 km s$^{-1}$ and +26 km s$^{-1}$, respectively, relative to the photospheric velocity.  These N I lines likely are interstellar.  For GD 40, the lines from Fe II at 1199.67 {\AA}, 1200.24 {\AA} and 1200.75 {\AA} dominate in the model spectrum, and there is no evidence for any photospheric nitrogen.  }
 \end{figure} 

 \begin{figure}
  \plotone{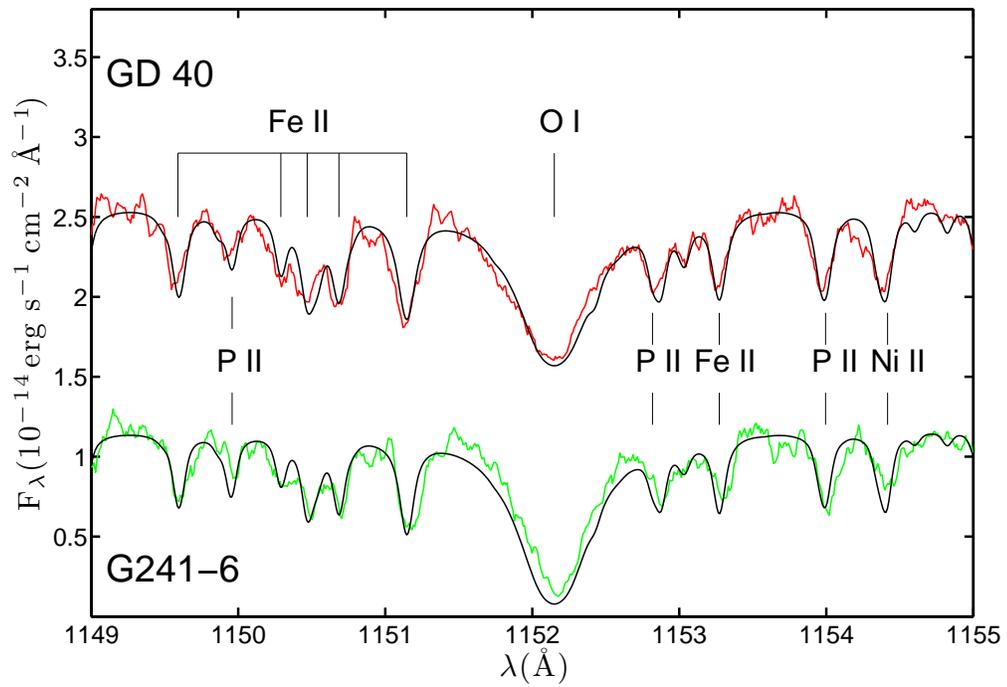}
 \caption{Similar to Figure 1 except for the O I line at 1152.15 {\AA} as well as lines of P, Fe and Ni.   }
 \end{figure}

 \begin{figure}
  \plotone{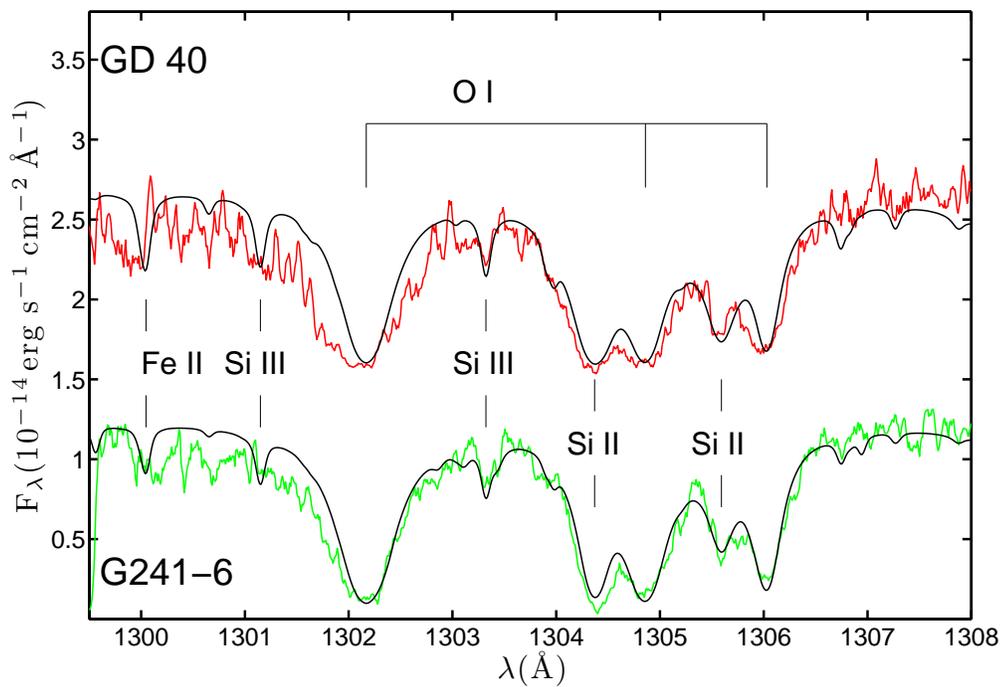}
 \caption{Similar to Figure 1 except for the O I lines at 1302.17 {\AA}, 1304.86 {\AA} and  1306.83 {\AA} as well as lines of  Fe and Si. These are ``night time" data and therefore
 we have less exposure time on-source and a lower signal-to-noise ratio in the data.  }
 \end{figure}

  \begin{figure}
  \plotone{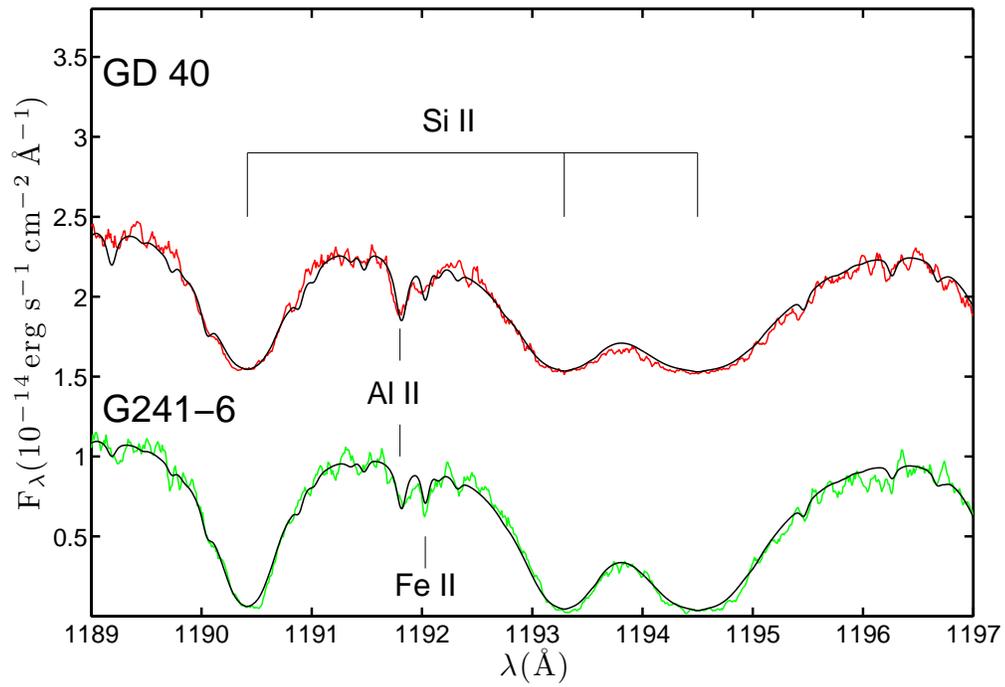}
 \caption{Similar to Figure 1 except for the spectral region near the Al II  line at 1191.8 {\AA}.  In GD 40, the aluminum line is stronger than the adjacent iron line; in G241-6, the aluminum line, if real, is
 weaker than the adjacent iron line.  Because iron is more abundant in GD 40 than in G241-6, these spectra directly demonstrate that aluminum is substantially more
 abundant in GD 40.   }
 \end{figure}
 
  \newpage
 
   \begin{figure}
  \plotone{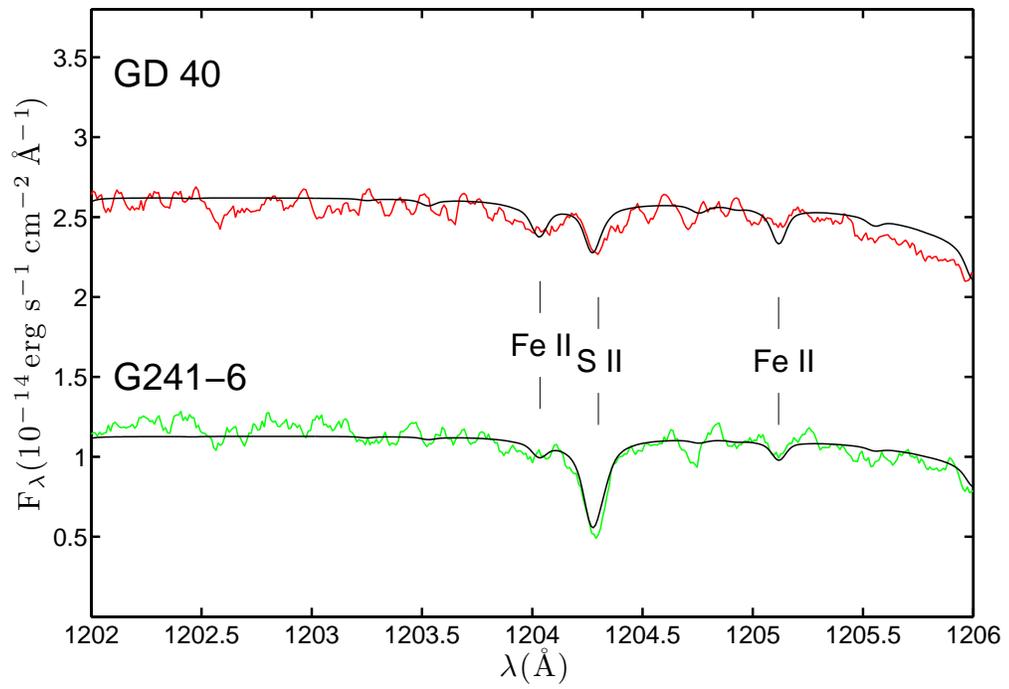}
 \caption{Similar to Figure 1 except for   the spectral region near the S II doublet at 1204.27 {\AA} and 1204.32 {\AA}. }
 \end{figure}
 
 \newpage
 
 \begin{figure}
 \plotone{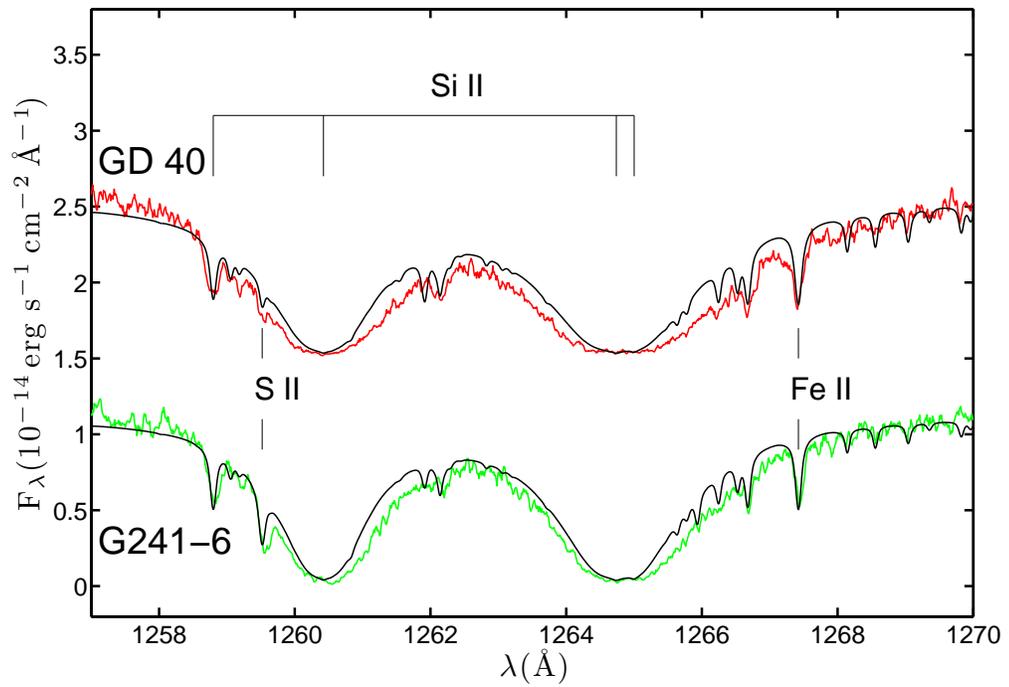}
 \caption{Similar to Figure 1 except for the spectral region near the Si II lines near 1260 {\AA}.}
 \end{figure}

  \begin{figure}
  \plotone{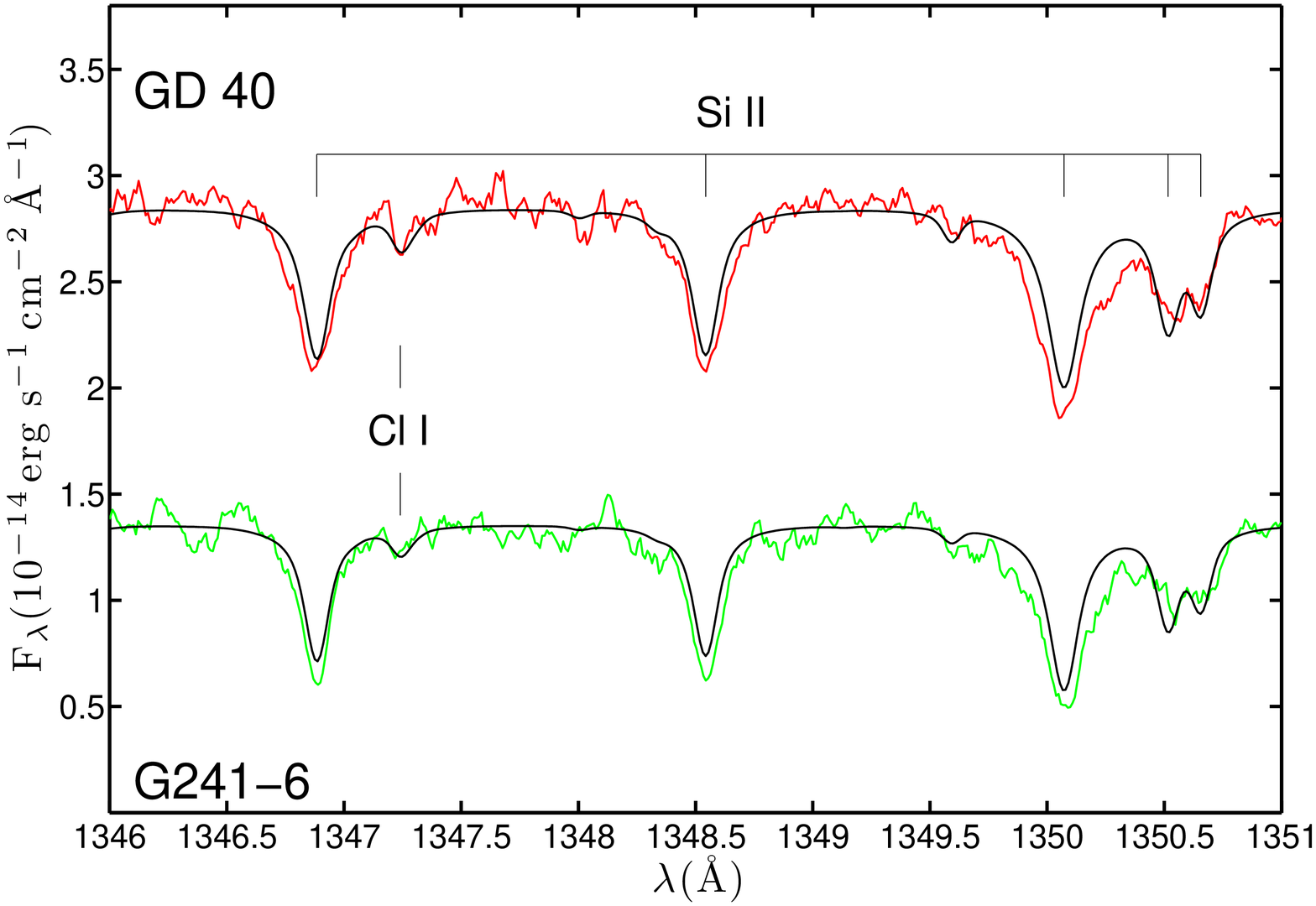}
 \caption{Similar to Figure 1 except for the spectral region near the undetected Cl I line at 1347.24 {\AA} }
 \end{figure}
 
 \begin{figure}
 \plotone{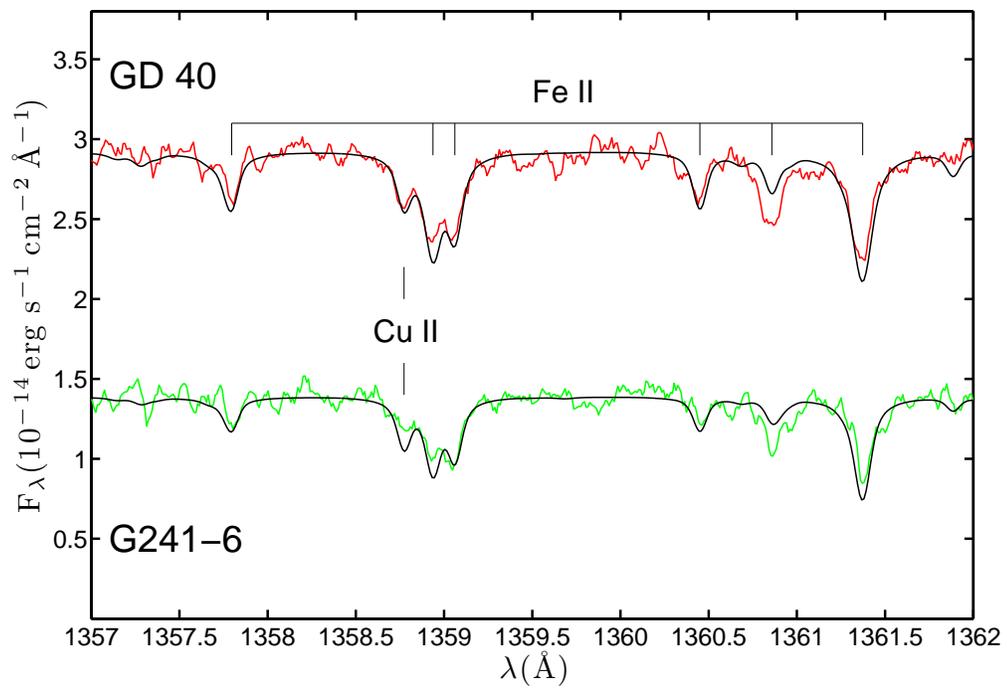}
 \caption{Similar to Figure 1 except for the spectral region near the undetected Cu II line at 1358.77 {\AA}. } 
  \end{figure}
 
 \newpage
 \begin{figure}
  \plotone{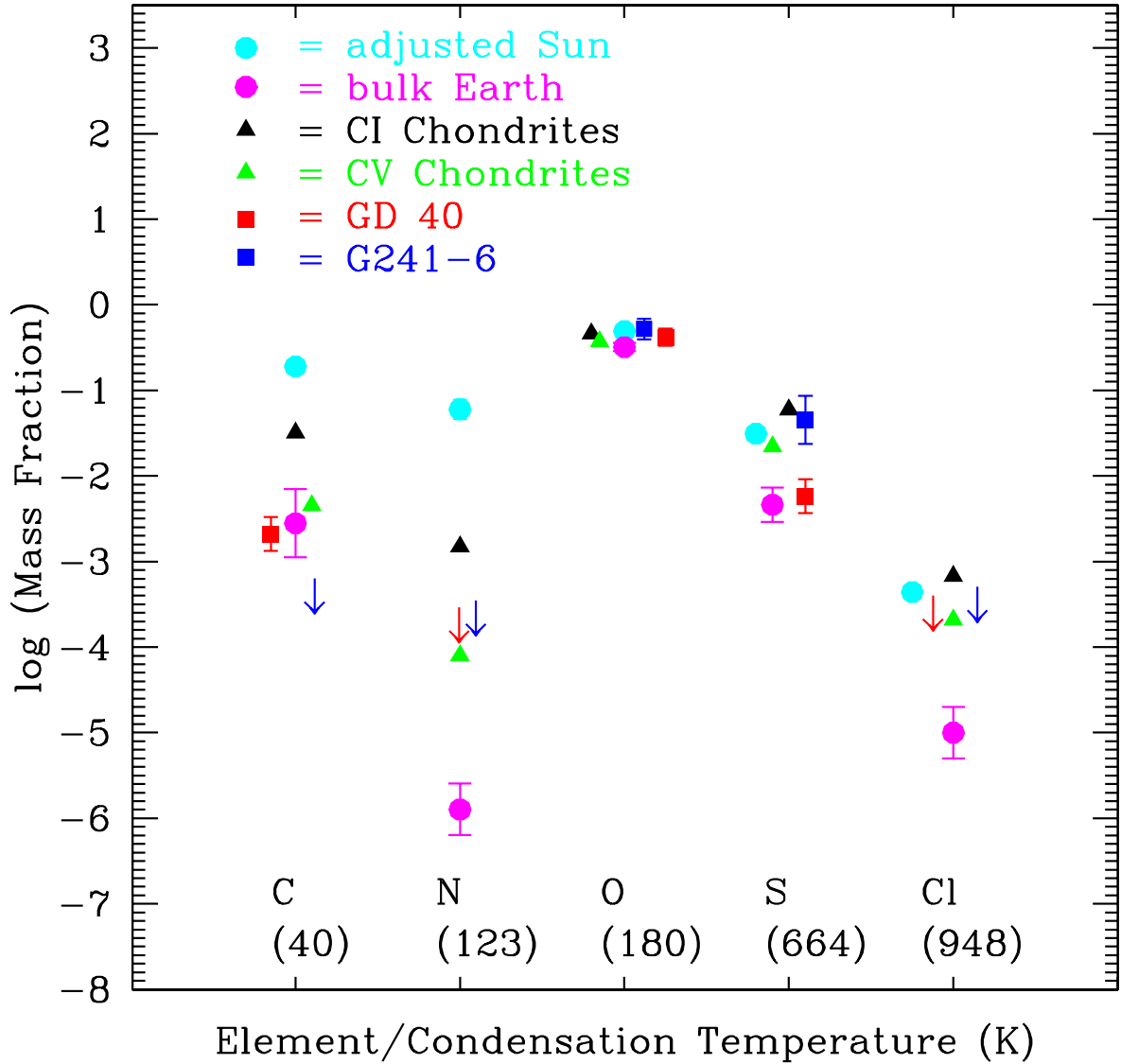}
 \caption{Display of results tabulated in Table 4: mass fractions of elements with 50\% condensation temperatures less than 1000 K  in the parent bodies accreted onto GD 40 and G241-6 and, for comparison,
 both CI and CV chondrites \citep{Wasson1988}, bulk Earth \citep{Allegre2001} and an ``adjusted Sun" which is all elements except hydrogen, helium and the noble gases given by \citet{Lodders2003}.  The temperature at which 50\% of the element condenses in the calculations of \citet{Lodders2003} is shown beneath its symbol on the plot.
 The relationship between the mass fractions of pollutants in the stellar atmosphere and the bulk composition of the accreted parent body are discussed in Section 4.  To make
 a conservative estimate of the mass fraction of the important volatiles -- carbon, nitrogen and oxygen -- we assume the system is in the build-up phase.  If the system is in
 a steady state or declining phase, then the results plotted here  overestimate the mass fractions of the lighter elements and underestimate the mass fractions of the heavier elements in the parent bodies.
 }
 \end{figure}
 \begin{figure}
  \plotone{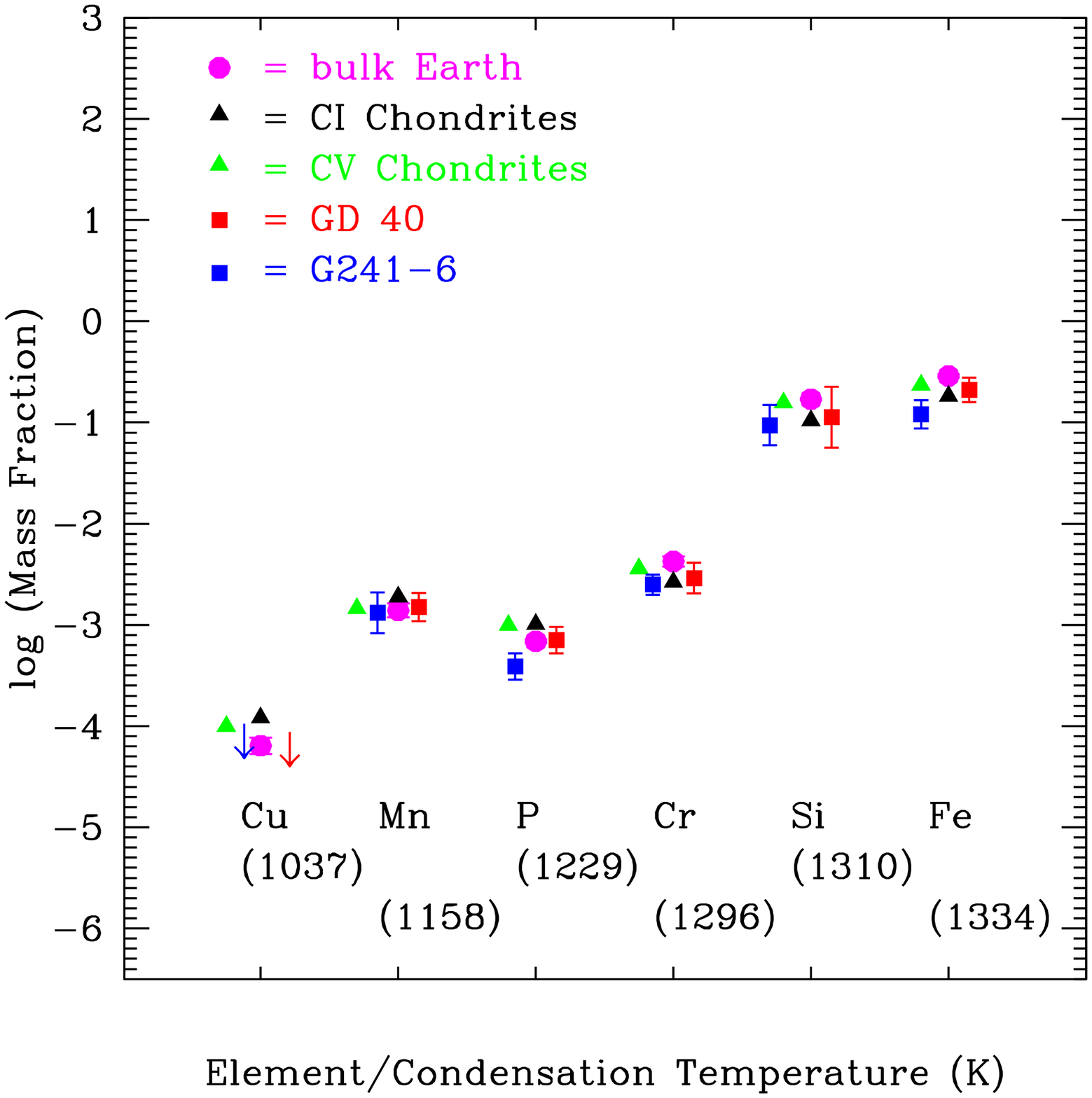}
 \caption{Similar to Figure 11 for elements with condensation temperatures greater than 1000 K and less than 1335 K.    We do not  display ``adjusted Sun" because it has  the same pattern as ``CI Chondrites". }
 \end{figure}
 
 \begin{figure}
  \plotone{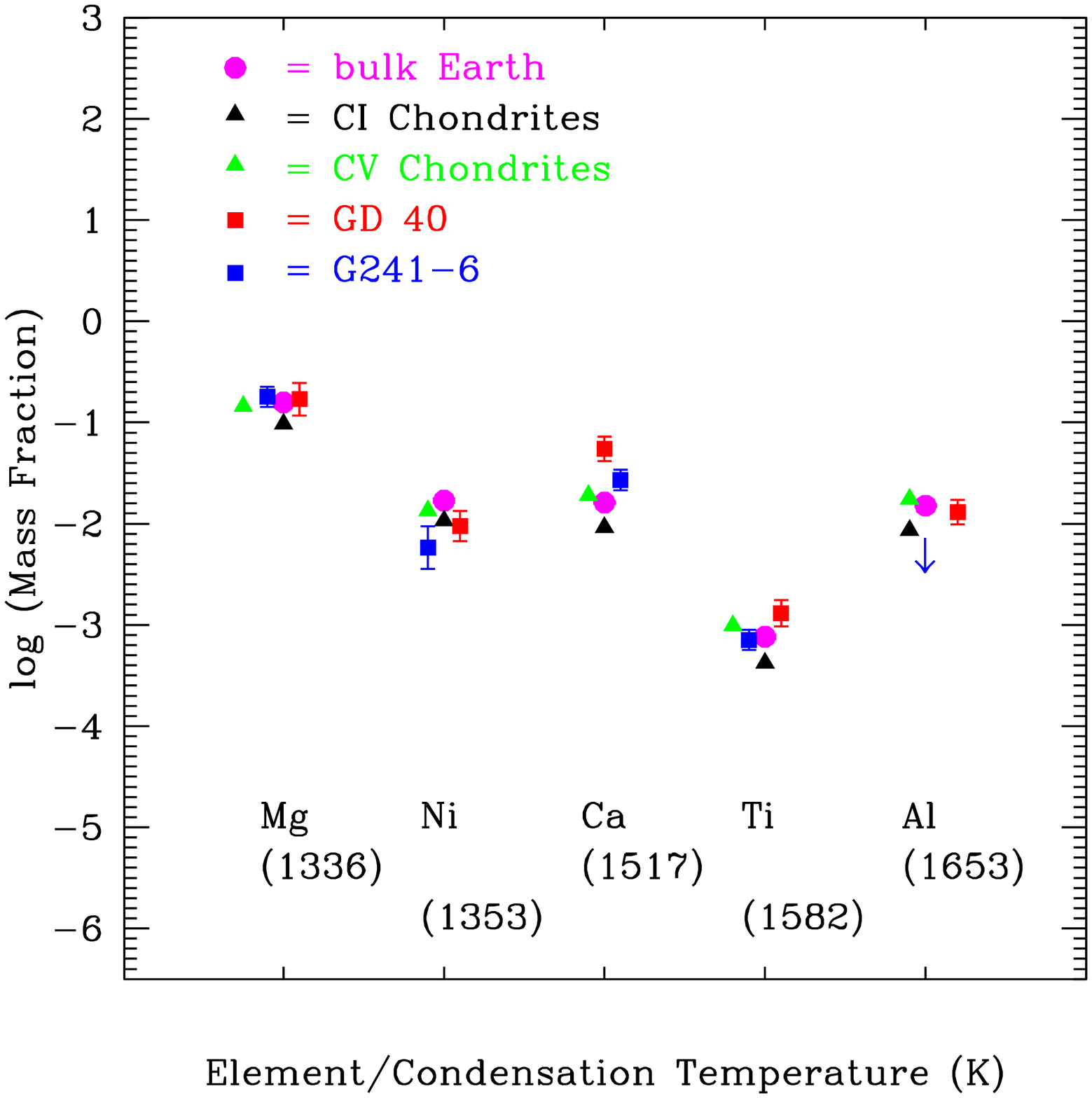}
 \caption{Similar to Figure 11 for elements with condensation temperatures greater than 1335 K.  We do not  display ``adjusted Sun" because it has  the same pattern as ``CI Chondrites".  }
 \end{figure}
\end{document}